 \documentclass[11pt]{article}

\usepackage[a4paper,margin=1in]{geometry}
\usepackage{graphicx}
\usepackage{float}
\usepackage{subcaption}
\usepackage{xparse}
\usepackage{pgffor}
\usepackage{amsmath,amssymb}
\usepackage{anyfontsize}

\graphicspath{{./figures/}}

\usepackage[
  colorlinks=true,
  linkcolor=blue,
  citecolor=blue,
  filecolor=blue,
  urlcolor=blue
]{hyperref}

\begin{document}

\title{Chaos Growth in a Generic 2D Hamiltonian System}
\author{
A.C. Tzemos\thanks{Corresponding author: atzemos@academyofathens.gr}, G.Contopoulos\thanks{email: gcontop@academyofathens.gr}\\[0.5em]
\small Research Center for Astronomy and Applied Mathematics,
Academy of Athens,\\ Soranou Efessiou 4, GR-11527 Athens, Greece}

\date{}
\maketitle

\begin{abstract}

We examine the geometric mechanism controlling the transition from local
to global chaos in a generic 2D Hamiltonian system, the generalized H\'{e}non-Heiles system. Using
Poincar\'e sections and invariant manifolds, we show that the chaotic
fraction grows through abrupt transitions associated with the
destruction of invariant barriers. We introduce an index that gives the time required for the  first penetration of a local chaotic domain into the global chaos. This index increases sharply and tends to infinity as
a critical energy is approached. The set of lobes generates heteroclinic connections, while multiple unstable periodic orbits exhibit a global manifold structure characterised by repeated loops and lobes. However, in special
chaotic cases close to the integrable H\'{e}non-Heiles potential, where
two asymptotic curves extend to infinity, the manifold
topology differs qualitatively from the generic case, as the
asymptotic curves no longer form closed structures surrounding the
periodic orbits.
\end{abstract}

\section{Introduction}

Chaos in dynamical systems is characterised by a sensitive dependence on initial conditions: even infinitesimal differences in the  starting points lead to exponentially diverging trajectories as time evolves, while the system is confined within a finite phase space \cite{lichtenberg2013regular,luo2012regularity}.

The concept of chaos can be traced back to the pioneering work of Poincaré (1899), who first recognized the intricate and unpredictable nature of certain dynamical systems. Lacking modern computational tools, he remarked on the overwhelming complexity of these structures: ``One is struck by the complexity of this figure, which I am not even attempting to draw'' \cite{poincare1899methodes}. His insights laid the groundwork for what would later become a central field of study in mathematics and physics.

The modern era of Hamiltonian chaos theory \cite{bountis2012complex} began with a paper by H\'{e}non and Heiles in 1964 \cite{henon1964applicability}, which provided one of the first numerical explorations of chaotic motion in a realistic physical system. Following this work, extensive research efforts were devoted to understanding the mechanisms and implications of chaos. A comprehensive overview of these developments with many references can be found in  \cite{Contopoulos200210}.

In integrable Hamiltonian dynamical systems, motion is regular and confined to invariant tori, which correspond to quasiperiodic trajectories on closed, torus-shaped manifolds in phase space, and  chaos does not arise, since the dynamics is governed by as many conserved quantities as degrees of freedom. However, when a small perturbation is introduced, the system typically becomes non-integrable, and the phase space acquires a mixed structure in which regions of regular motion coexist with regions of chaotic dynamics. In this regime, chaotic behaviour usually appears first in the vicinity of unstable periodic orbits, where the local dynamics is organised by stable and unstable asymptotic manifolds, which act as skeletons for the surrounding phase space structure.

Nevertheless, according to KAM theory \cite{kolmogorov1954preservation,arnold1963proof,moser1962invariant}, a significant measure of invariant tori survives for sufficiently small perturbations, so that regular motion persists on large subsets of phase space and chaos remains confined to thin layers surrounding resonances.

As the perturbation strength increases, these chaotic layers gradually widen and begin to interact, while invariant tori are progressively deformed and destroyed, leading to increasingly efficient transport across phase space by chaotic orbits. Eventually, beyond a certain threshold, the growth of chaotic domains becomes abrupt, a transition first observed numerically by H\'{e}non and Heiles \cite{henon1964applicability}. This transition is due to the overlap of nonlinear resonance zones \cite{Rosenbluth1966,Contopoulos1966c} which eliminates the remaining invariant barriers and allows for the emergence of large-scale global chaos driven by the interaction and intersection of asymptotic curves associated with different unstable periodic orbits.

In systems with two degrees of freedom much work was done in 2D Hamiltonians of the generalized H\'{e}non--Heiles form
\begin{equation}\label{Ham}
H = H_0 + \varepsilon H_1 = h,
\end{equation}
where
\begin{equation}
H_0 = \frac{1}{2}\left(p_x^{2}+p_y^{2}+\omega_1^{2}x^{2}+\omega_2^{2}y^{2}\right),
\end{equation}
and
\begin{equation}\label{H1}
H_1 = xy^{2} + \alpha x^{3}.
\end{equation}
The original H\'{e}non-Heiles system has $\omega_1=\omega_2=1$, $\alpha=-\tfrac{1}{3}$, $h=1$ and various values of $\varepsilon$. It was realized that chaos appears near the unstable periodic orbits. In fact the unstable periodic orbits generate two stable and two unstable asymptotic curves, and chaos is produced close to them, when the asymptotic curves of different periodic orbits intersect.

Various aspects of Hamiltonian chaos have been studied in recent years (e.g. \cite{PhysRevLett.98.234301, blesa2012escape, di2021hamiltonian,salasnich2026geometric}). However, there are still some problems that have not been answered in sufficient detail up to now.

{Previous studies have established the roles of cantori, stickiness, homoclinic and heteroclinic tangles, and asymptotic-curve lobes in the organisation of chaotic motion \cite{mackay1987resonances,polymilis2003homoclinic,Contopoulos200210,shirts1982approximate,zaslavsky2002chaos}. In this work, we address the following questions concerning the transition from local to global chaotic transport:}

\begin{enumerate}
\item {How does the chaotic fraction evolve as the energy increases? We show that the chaotic fraction increases in a stepwise manner.}

\item {How does the geometry of invariant manifolds control the penetration of local chaotic domains into the global chaotic sea? In particular, we consider the stable and unstable asymptotic curves associated with simple and triple unstable periodic orbits and use the first-penetration lobe index to characterize this transport process. We quantify this process using the stable and unstable asymptotic curves.}

\item {How does the topology of the invariant manifolds change beyond the escape threshold? We show that when two asymptotic branches extend to infinity, the topology becomes qualitatively different.}

\end{enumerate}


{The structure of the paper is as follows. Section 2 characterises the stepwise growth of the chaotic fraction with energy and identifies the invariant-curve destructions associated with the appearance of new routes to chaos. Section 3 introduces and quantifies the first-penetration lobe index for the transport of asymptotic curves from simple and triple unstable periodic orbits into the global chaotic domain, and examines the associated heteroclinic connections and high-period periodic orbits. Section 4 investigates the qualitatively different topology that emerges when two asymptotic branches extend to infinity. Finally, Section 5 summarises our results. In the Appendix we provide  details about the accuracy of the calculations of the invariant manifolds.}

\section{{Stepwise Growth of the Chaotic Fraction with Energy}}

{We first examine how the measure of the chaotic region changes with energy. The resulting $P_{ch}(h)$, which  is defined as the total area of the
chaotic domains divided by the total area of the Poincar\'e section,
$2\pi h/\omega_1$, provides a global signature of the successive changes in the invariant-curve structure. In particular, $P_{ch}(h)$ grows through a sequence of abrupt transitions associated with the destruction of invariant curves surrounding resonant island chains.}

In a 2-D conservative problem, like the Hamiltonian (1)--(3), the basic characteristics of the orbits are found if we consider the intersections of these orbits with the plane $y=0$, for fixed values of $\varepsilon$ and $h$ (Poincar\'{e} sections). We consider the successive points $(x,p_x)$, with $p_y>0$, while $p_y$ is obtained from Eq.~\eqref{Ham}. The orbits are contained inside a limiting curve 
\begin{equation}\label{lc}
p_x^2=2h-\omega_1^2x^2-2\varepsilon \alpha x^3=0.
\end{equation}
Outside this curve there are no real orbits because $p_x^2<0$.

{In the following figures we consider the case with $\alpha=0$, $\varepsilon=0.08$, $\omega_1=\sqrt{1.6}$, $\omega_2=\sqrt{0.9}$, and increasing values of $h$ (in order to compare our results with those of our previous papers  \footnote{In some papers we increase the value of $\varepsilon$ keeping $h$ constant. But the two cases are equivalent because $\varepsilon \sqrt{h}=\varepsilon_0\sqrt{h_0}$.}). This is a resonant case with $\omega_1/\omega_2=4/3$.}

{The system contains a central stable periodic orbit $O$ and a triple
periodic orbit $(O_1O_2O_3)$, which is unstable and lies between the
three surrounding stable period-three islands. The chaotic domain is
initially concentrated around this unstable orbit and is progressively
enlarged as the energy increases and the invariant curves surrounding
the resonant island chains are destroyed.}

\begin{figure}[H]
\centering
\includegraphics[scale=0.2]{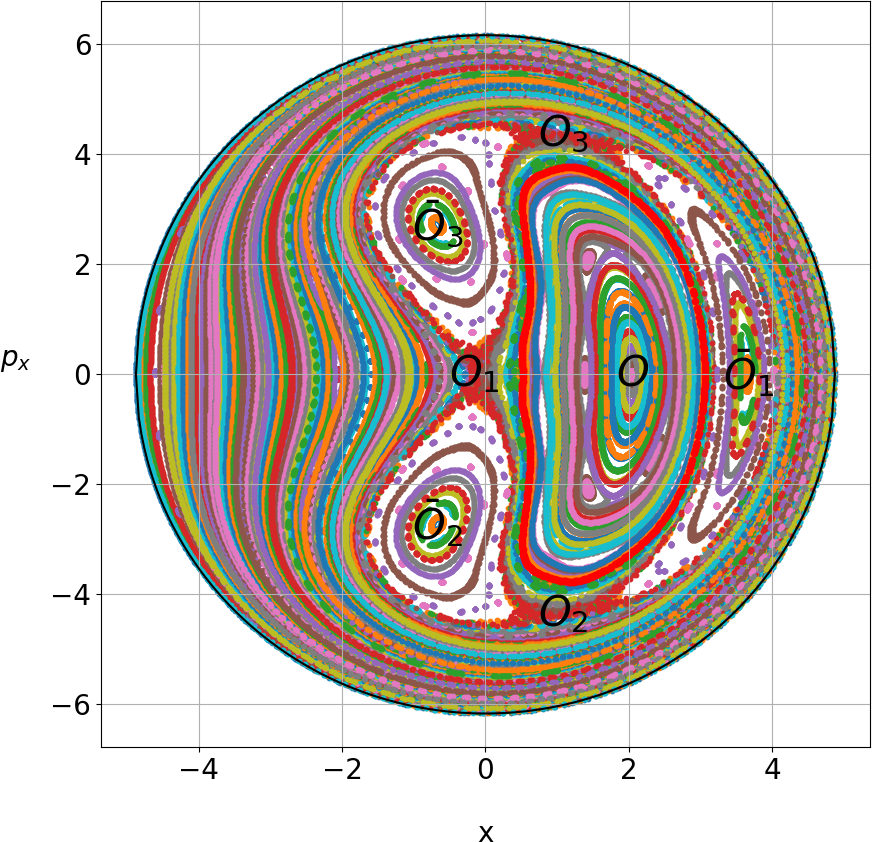}[a]
\includegraphics[scale=0.2]{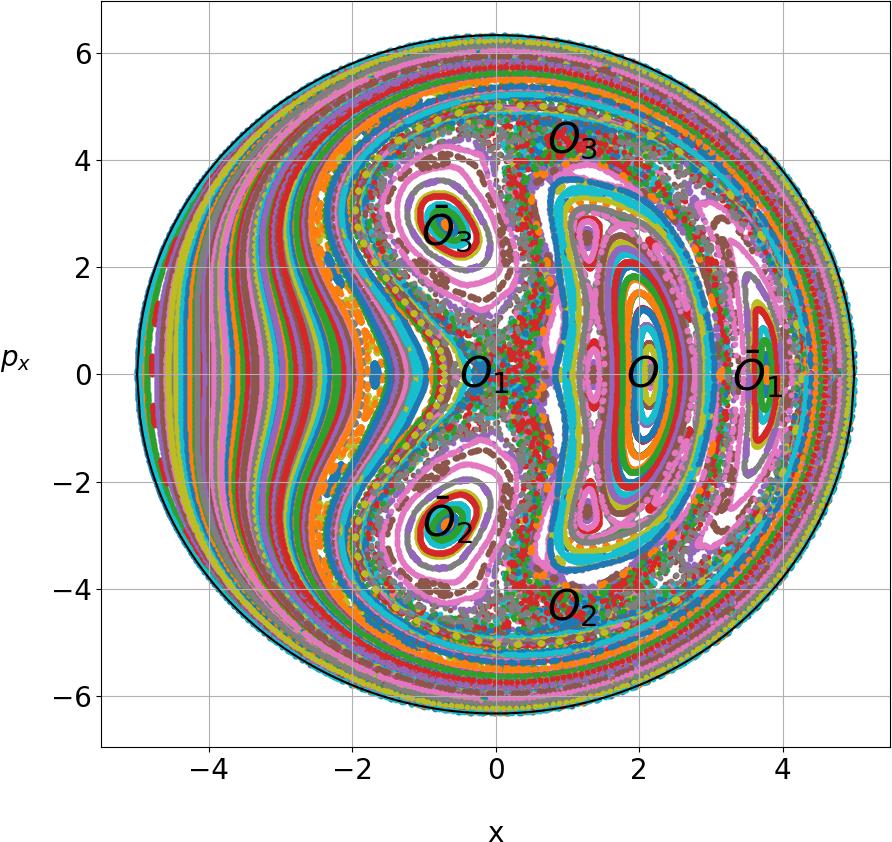}[b]\\
\includegraphics[scale=0.2]{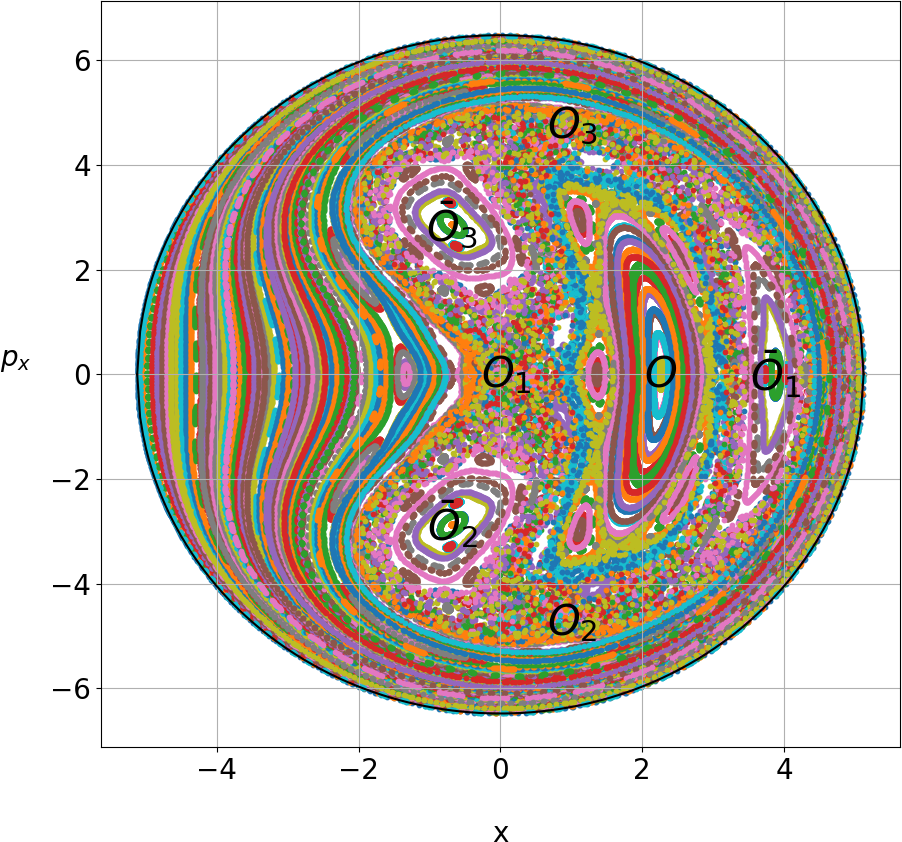}[c]
\includegraphics[scale=0.2]{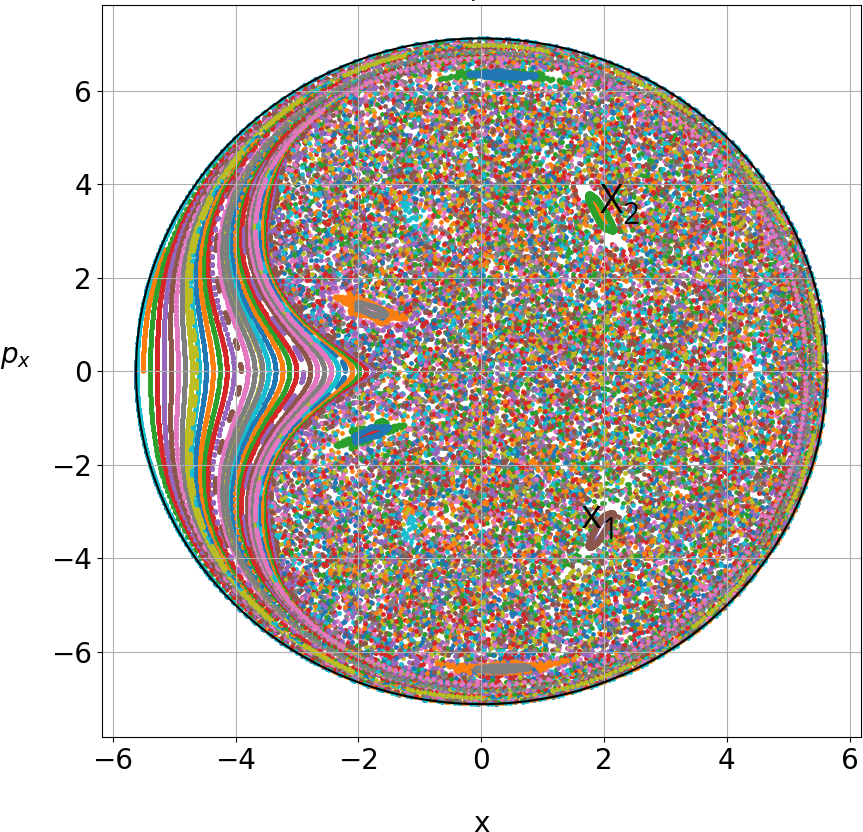}[d]
\caption{{Evolution of the Poincaré surface of section $(y=0,\;p_y>0)$ with increasing energy for $\alpha=0$, $\varepsilon=0.08$, $\omega_1=\sqrt{1.6}$, and $\omega_2=\sqrt{0.9}$: (a) $h=19$, (b) $h=20$, (c) $h=21$, and (d) $h=25.3$. The successive destruction of invariant curves surrounding resonant island chains leads to the progressive enlargement and connection of chaotic domains. At $h=25.3$, the phase space is predominantly chaotic, while small regular regions remain near the boundary.}}\label{fig1} \end{figure}

{At $h=19$ (Fig.~\ref{fig1}(a)), the chaotic domains remain relatively
small and are separated by invariant curves. As the energy increases to
$h=20$ (Fig.~\ref{fig1}(b)), the invariant curve surrounding the
eight-island chain is destroyed and replaced by a cantorus, allowing the
surrounding chaotic domain to communicate with smaller chaotic regions.
At $h=21$ (Fig.~\ref{fig1}(c)), the last invariant curve surrounding the
five-island chain is destroyed, producing a further enlargement of the
chaotic domain. By $h=25.3$ (Fig.~\ref{fig1}(d)), the phase space is
predominantly chaotic, although regular regions and invariant curves
remain near the boundary. This value lies just below the escape energy,
$h_{\rm esc}=25.3125$. Higher-order stable and unstable periodic orbits and their associated
islands are also present around the central orbit $O$, as seen in
Fig.~(\ref{fig1}a).}


{As $h$ increases, successive invariant curves surrounding resonant
island chains are destroyed. The resulting cantori allow previously
separated chaotic domains to communicate, producing abrupt increases in
the total chaotic fraction. The transitions illustrated in Fig.~\ref{fig1}
therefore provide the local phase-space picture underlying the stepwise
growth of $P_{\rm ch}(h)$.}



{For larger $h$, the phase space becomes predominantly chaotic, with only
small regular regions and invariant curves remaining near the boundary
(Fig.~\ref{fig1}(d)). At $
h_{\rm esc}=25.3125$\footnote{{In fact the curve of zero velocity is given by Eqs.~\eqref{Ham}-\eqref{H1} as
\begin{equation}
y^2=\frac{2h-\omega_1^2x^2}{\omega_2^2+2\varepsilon x}
\end{equation}
and this goes to infinity for  $x=x_c=-\frac{\omega_2^2}{2\varepsilon}=-5.625$ and $2h-\omega_1^2x^2>0$. The escape energy appears when $h=\omega_1^2x_c^2/2=25.3125$. The escapes occur along the $y$ direction.}}, the limiting curve opens and escape becomes possible along the $y$
direction. The topology of the invariant manifolds beyond this threshold
is examined separately in Section~4.}

\begin{figure}[H]
\centering
\includegraphics[scale=0.17]{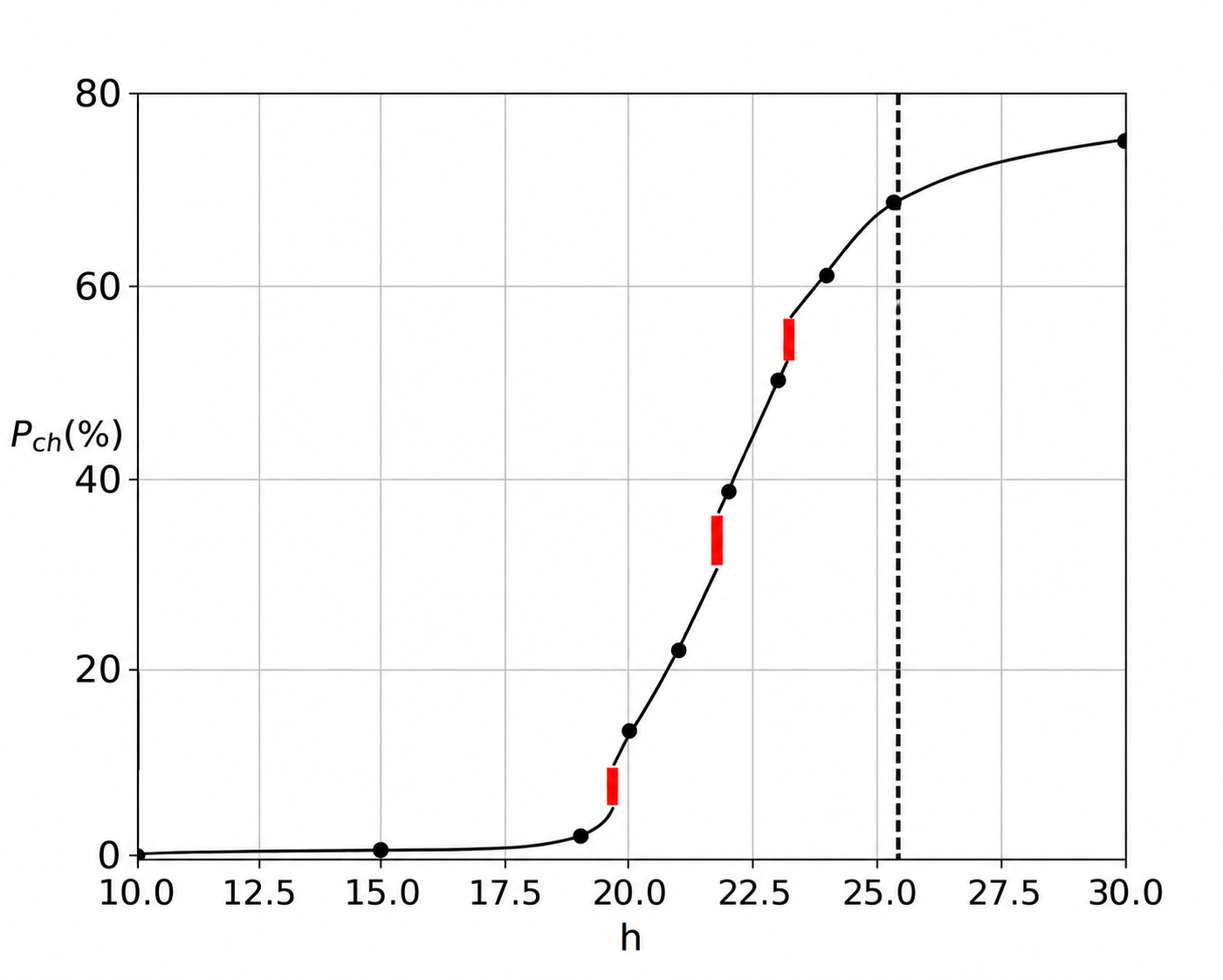}
\caption{The proportion $P_{ch}$ of chaotic areas as a function of the energy $h$. Beyond the escape energy $h=h_{esc}=25.3$  (vertical line) $P_{ch}$ includes the escaping orbits. We observe the abrupt jumps of the chaos growth (red segments).}\label{area}
\end{figure}

{Figure~\ref{area} shows the proportion $P_{\rm ch}$ of chaotic areas as a
function of the energy $h$. The chaotic fraction remains small up to
$h\simeq19$ and then increases rapidly, reaching values above $60\%$
near the escape energy. Although the overall trend appears smooth,
$P_{\rm ch}(h)$ exhibits a sequence of abrupt increases associated with
the successive destruction of invariant barriers. The most pronounced
transition occurs near $h\simeq23.3$, where the periodic orbit $O$
becomes unstable and the last invariant curve surrounding the resulting
local chaotic domain is destroyed, allowing this domain to be connected with
the outer chaotic sea.}
 
{The stepwise behaviour of $P_{\rm ch}(h)$ provides a global
signature of the underlying transport transitions. The formation of
cantori following the destruction of invariant curves is a known
mechanism for the connection of chaotic domains
\cite{mackay1987resonances,Contopoulos200210}. The central question for
the present study is therefore how efficiently the corresponding
invariant manifolds penetrate the newly connected global chaotic
region. We address this quantitatively in the next section by calculating
the lobe order required for the first penetration into the global
chaotic sea.}
 

{Beyond the escape energy, $P_{\rm ch}$ also includes escaping orbits,
whose fraction increases smoothly with $h$ (in fact for $h=50$ the proportion of escaping orbits reaches about $98\%$).}

\section{{Approach to Global Chaos}}

{The central dynamical question is how the local chaotic structure generated near an unstable periodic orbit communicates with the pre-existing global chaotic sea. This is a transport problem controlled by the geometry of the stable and unstable asymptotic curves. To quantify it, we follow the lobe hierarchy generated by the invariant manifolds and identify the first lobe that penetrates the global chaotic domain.}

We compute the asymptotic curves from the eigenvectors by successive iterations up to order \(N\), which determines the development of the manifold geometry in the Poincaré map.
 {We additionally introduce the first-penetration lobe index $i(h)$, defined as the index of the first outer lobe $U_i'$ whose extent reaches the global chaotic sea. This index provides a discrete measure of the transport  constraint imposed by the remaining invariant structures.} Because of the symmetry of the system with respect to the variable $y$, we calculate only one stable and one unstable branch, starting downwards. In the following subsections we study the asymptotic manifolds associated with the simple unstable periodic orbit $O$ and with the triple unstable periodic orbit $O_1O_2O_3$.

\subsection{The case of the simple unstable periodic orbit $O$}\label{31}

\begin{figure}[H]
\centering
\includegraphics[scale=0.25]{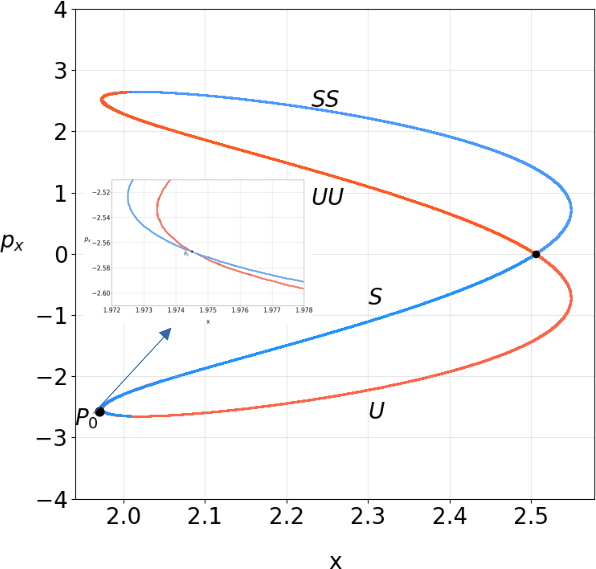}[a]
\includegraphics[scale=0.22]{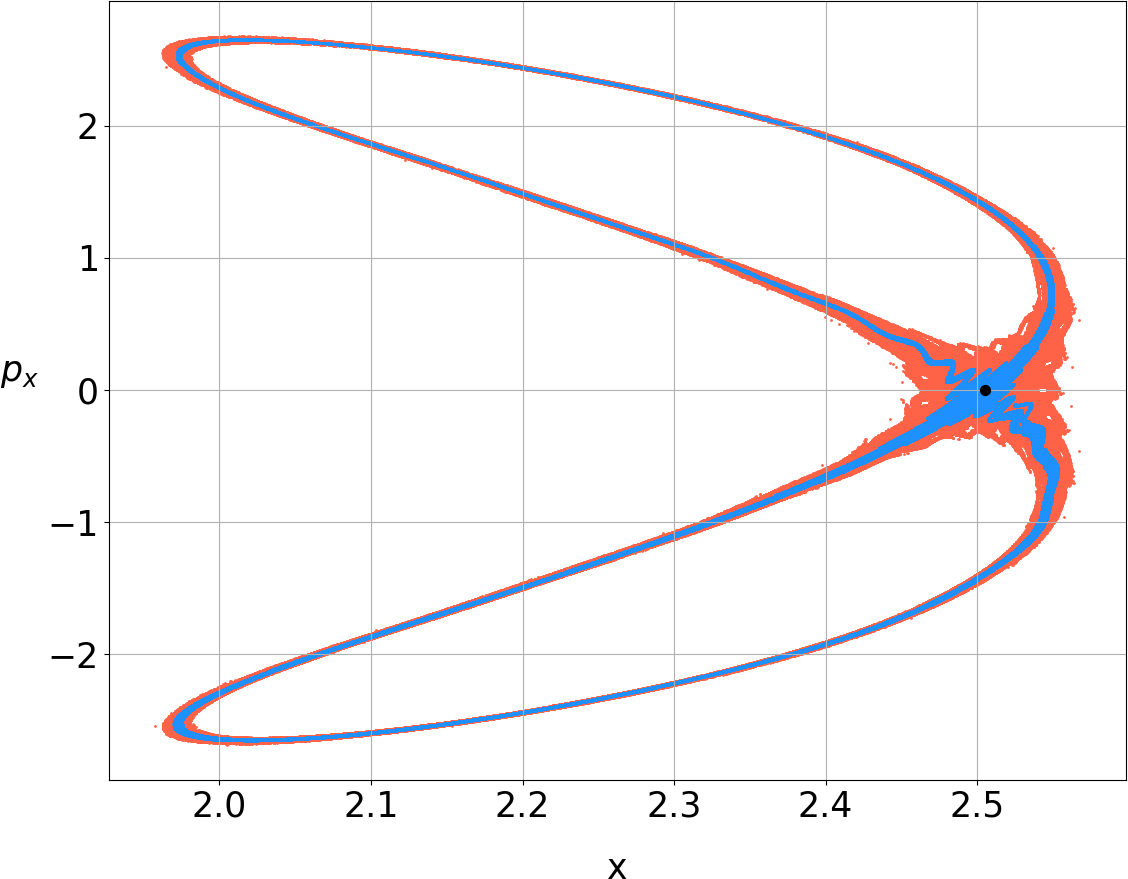}[b]\\
\includegraphics[scale=0.22]{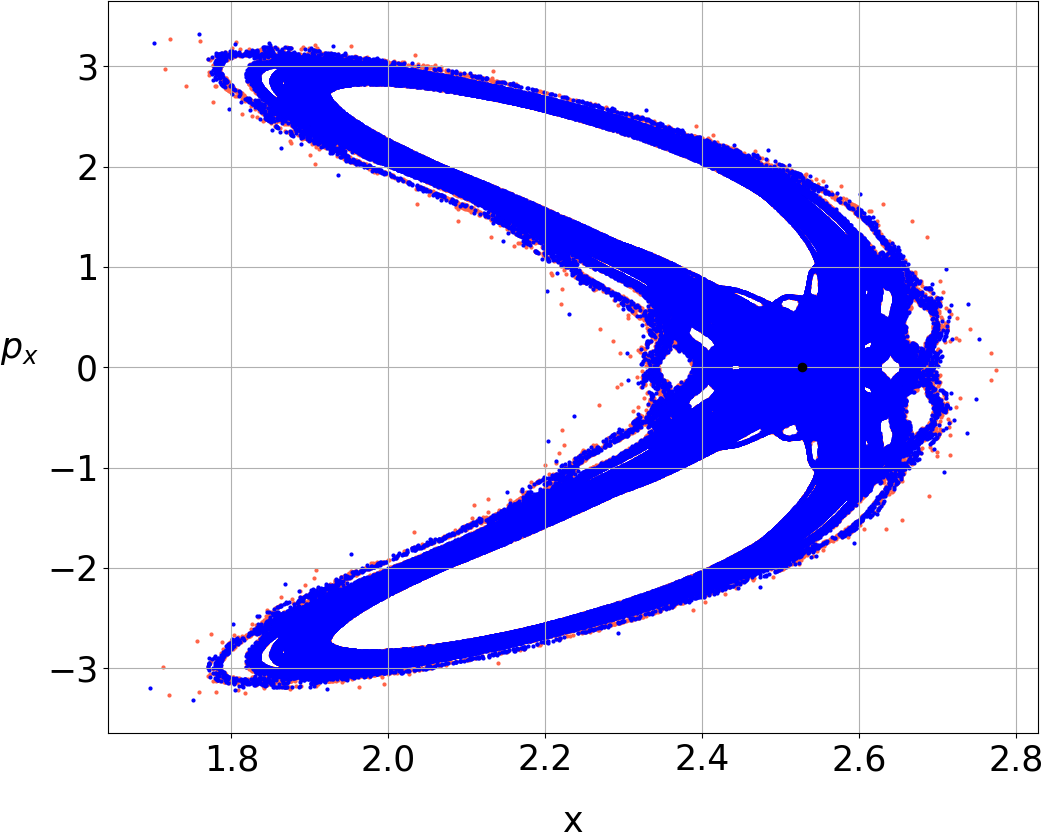}[c]
\includegraphics[scale=0.22]{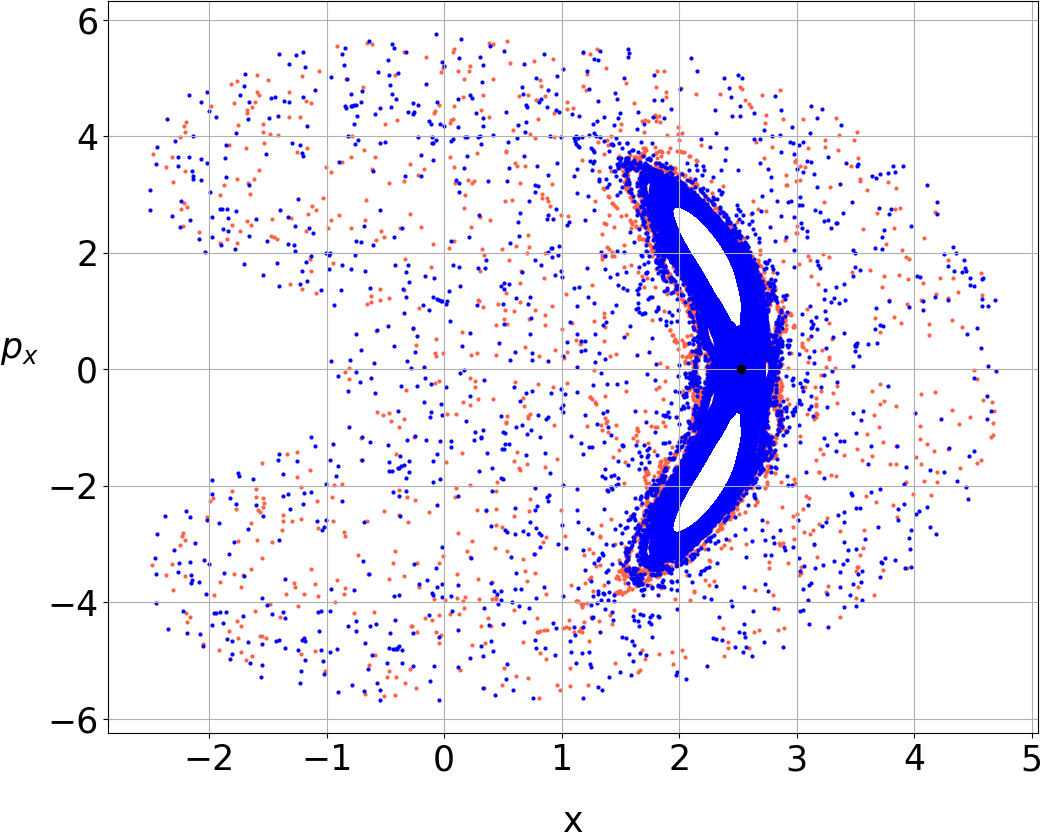}[d]
\caption{The asymptotic curves of the central unstable periodic orbit $O$ in the case  $h=23.2$ for a) $N=50$  and b) $N=700$ iterations and in the case $h=23.4$ for c) $N=444$ and d) $N=800$ iterations (20000 initial conditions along every eigenvector starting at distance $\delta s_0=10^{-7}$ from the periodic orbit, up to a distance $\delta s = 10^{-6}$). }\label{mans}
\end{figure}

In Fig.~(\ref{mans})a   we calculate the form of the asymptotic  curves $U$ and $UU$ (unstable) and $S$ and $SS$ (stable) of the periodic orbit $O$ ($x=2.5054 $, $p_x=0$) for $h=23.2$ up to $N=50$ iterations. The curves $U$ and $S$  intersect at a homoclinic point  $P_0$ $(x=1.9745,p_x=-2.567)$. In the  zoom of Fig.~3a we see the introduction of the curves $U$ and $S$ in the  region around $P_0$. A symmetric intersection point is between the curves $UU$ and $SS$.

Then in Figs.~\ref{mans}b we compute the  asymptotic curves for $h=23.2$ up to $N=700$ iterations. Their oscillations are so close and dense to each other that we can see only the overall area that they cover (the red and blue curves overlap). In the case $h=23.2$ the  chaotic  area is limited as $N$ increases because the overall chaotic region is surrounded by an invariant curve (Fig.~\ref{fig1}c). On the other hand in the case $h=23.4$ the extent of the chaotic domain increases gradually away from $O$ (for $y=0$) (Figs.~\ref{mans}c,d). This is consistent with the crossing of the cantorus formed by the destruction of the last KAM curve surrounding the region of $O$, including the upper and lower stable periodic orbits (Fig.~\ref{mans}). However, the time (number of iterations) needed to enter the large chaotic sea generated by the asymptotic curves of the triple periodic orbit $(O_1,O_2,O_3)$ is very long. This is an indication of the stickiness  of the region around cantori, a phenomenon which has been discussed in  \cite{shirts1982approximate, zaslavsky2002chaos}. In this case the orbit $O$ is unstable and below and above it we find the stable periodic orbits $X_1$ and $X_2$.

\begin{figure}[H]
\centering
\includegraphics[scale=0.12]{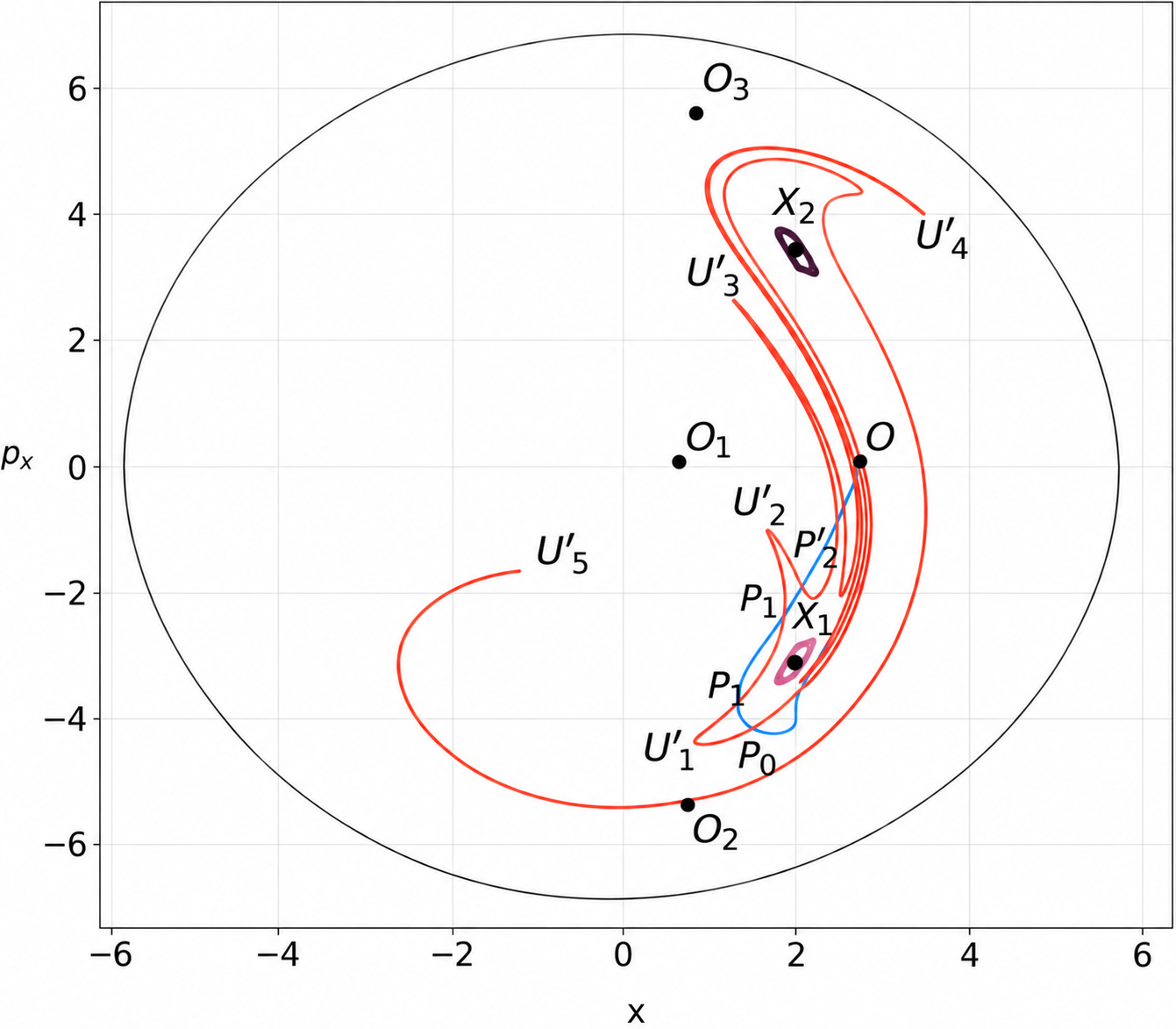}[a]
\includegraphics[scale=0.12]{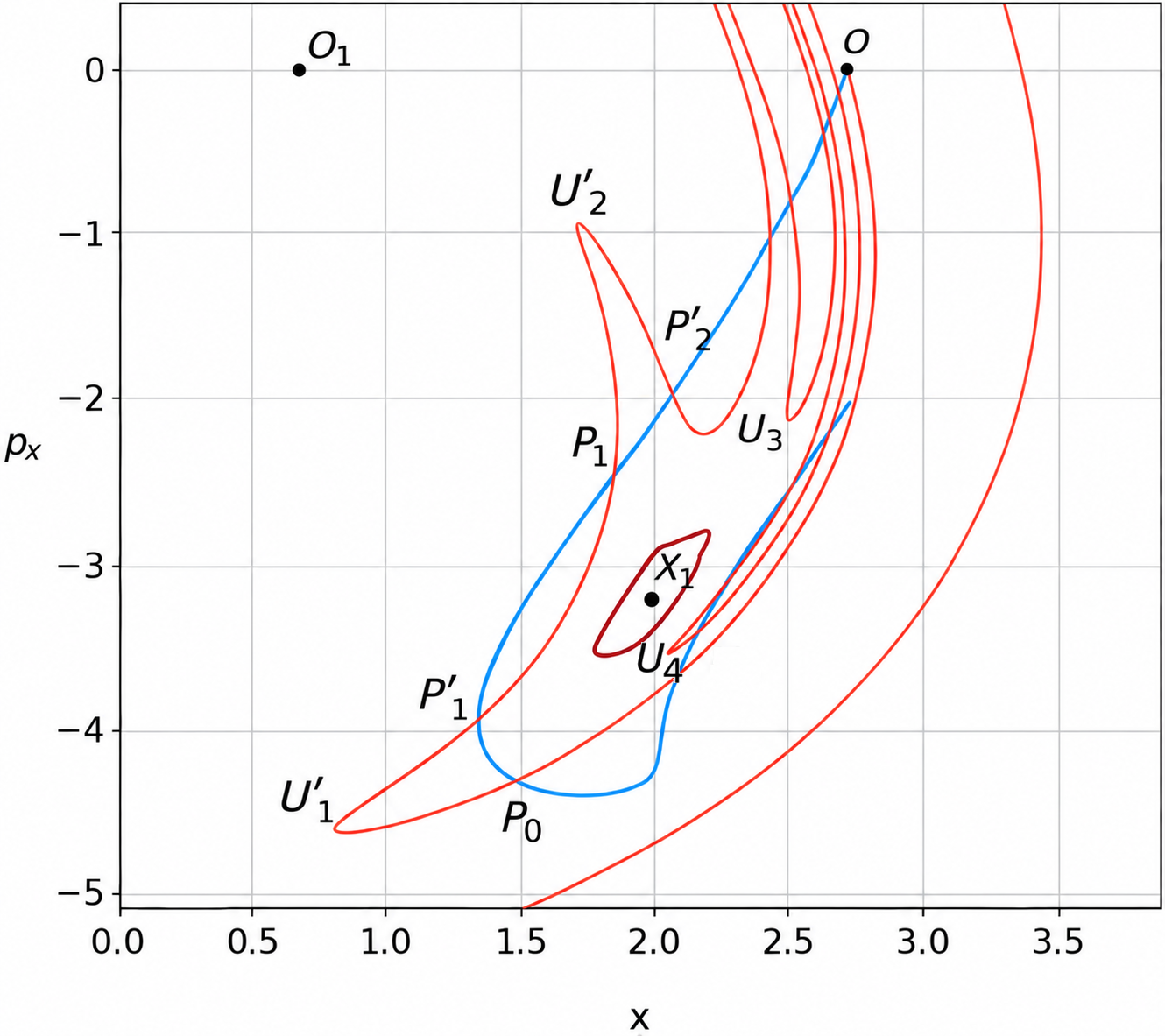}[b]
\includegraphics[scale=0.11]{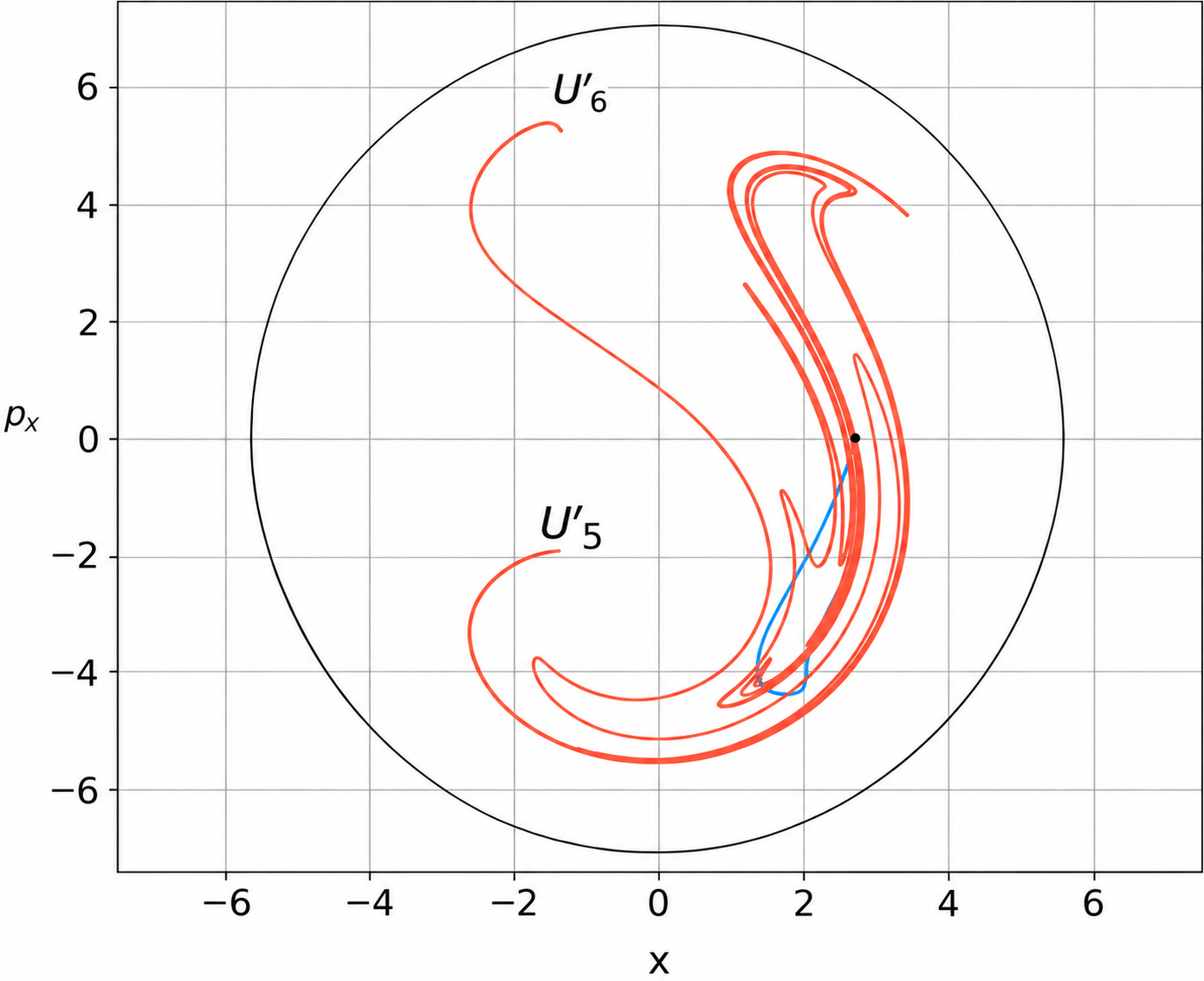}[c]
\includegraphics[scale=0.21]{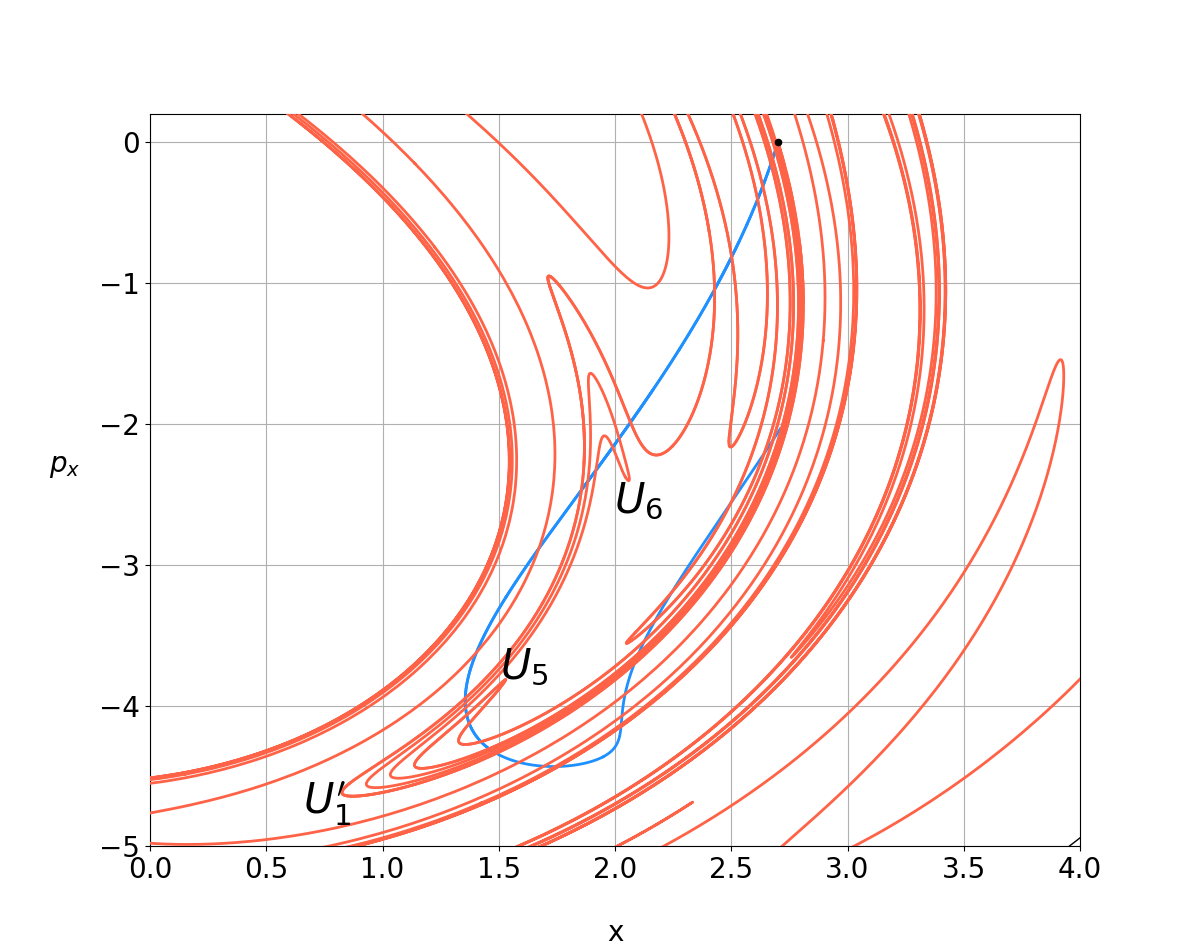}[d]
\caption{a) The lower asymptotic curves, $U$ (unstable, red) and $S$ (stable, blue) from $O$ intersect at the points $P_0,P_1',P_1,P_2',P_2\dots$ and form the lobes $U_1',U_2'\dots U_5'$ (outer) and $U_1,U_2,U_3,U_4$ (inner). The inner lobes are seen in more detail in b). c) The previous lobes plus the outer lobe $U_6'$ (and the inner lobes $U_5, U_6$ in the zoom (d)).  }\label{mans2}
\end{figure}

As $h$ increases, the time required for the asymptotic curves to enter the large chaotic sea decreases. This is seen  in Figs.~\ref{mans2}a,b (for $h\approx 25$).

In Fig.~\ref{mans2}a we show the unstable asymptotic curve $U$ for 20  iterations starting downwards from $O$ up to 20 iterations of the original segment and a relatively small part of the stable curve $S$. The two curves intersect at the homoclinic points $P_0, P_1', P_1,P_2',P_2...$ approaching the periodic orbit $O$. In fact, the curves $U$ and $S$ cannot intersect themselves but they intersect each other at all the homoclinic points.

The part of the curve $U$ between $P_0$ and $P_1'$ is called an outer lobe (lobe $U_1'$). The part of $U$ between $P_1'$ and $P_1$ is called an inner lobe (lobe $U_1$). In general the outer lobes are primed ($U_1',U_2'\dots$) and the inner lobes are unprimed ($U_1, U_2\dots$). Between the points $P_i$ and $P_{i+1}'$ we have an outer lobe ($U_i'$ from $P_i$ to $P_{i+1}$) and an inner lobe ($U_{i+1}$ from $P_{i+1}'$ to $P_{i+1}'$).

 The outer lobe $U_4'$ is the first that surrounds the islands above and below $O$ (both islands $X_1$ and $X_2$). In Fig.~\ref{mans2}a, we mark the tips of the outer lobes $U_1', U_2', U_3', U_4'$ and $U_5'$. We see that these lobes  become thinner as the order increases (Fig.~\ref{mans2}b is a zoom of the lower part of Fig.~\ref{mans2}a).


 The lobe $U'_5$ is the first lobe that goes far to the left, beyond the point $\bar{O}_2$, into the large chaotic sea due to the asymptotic curves of the orbit $O_1O_2O_3$.

Then in Fig.~\ref{mans2}c we have the previous lines, plus the outer lobe $U'_6$ and the inner lobes $U_5, U_6$. The inner lobe $U_5$ is the first inner lobe that enters  $U'_1$ (see the zoom of Fig~\ref{mans2}d). The  inner lobes $U_6,U_7,\ldots$ enter the outer lobes $U'_2,U'_3,\ldots$, while  $U'_6$ reaches the region around the point $\bar{O}_3$ (symmetric to $\bar{O}_2$). The main peculiarity of  $U'_6$ is that it makes some extra oscillations before reaching its tip. It starts at a point $P_5$ on the $S$-curve close to O and terminates at $P_6'$ closer to $O$ on $S$. Initially it is close to the curve $U_5'$ but later it deviates and makes some oscillations before reaching the tip at  the point $U_6'$.



\begin{figure}[H]
\centering
\includegraphics[scale=0.30]{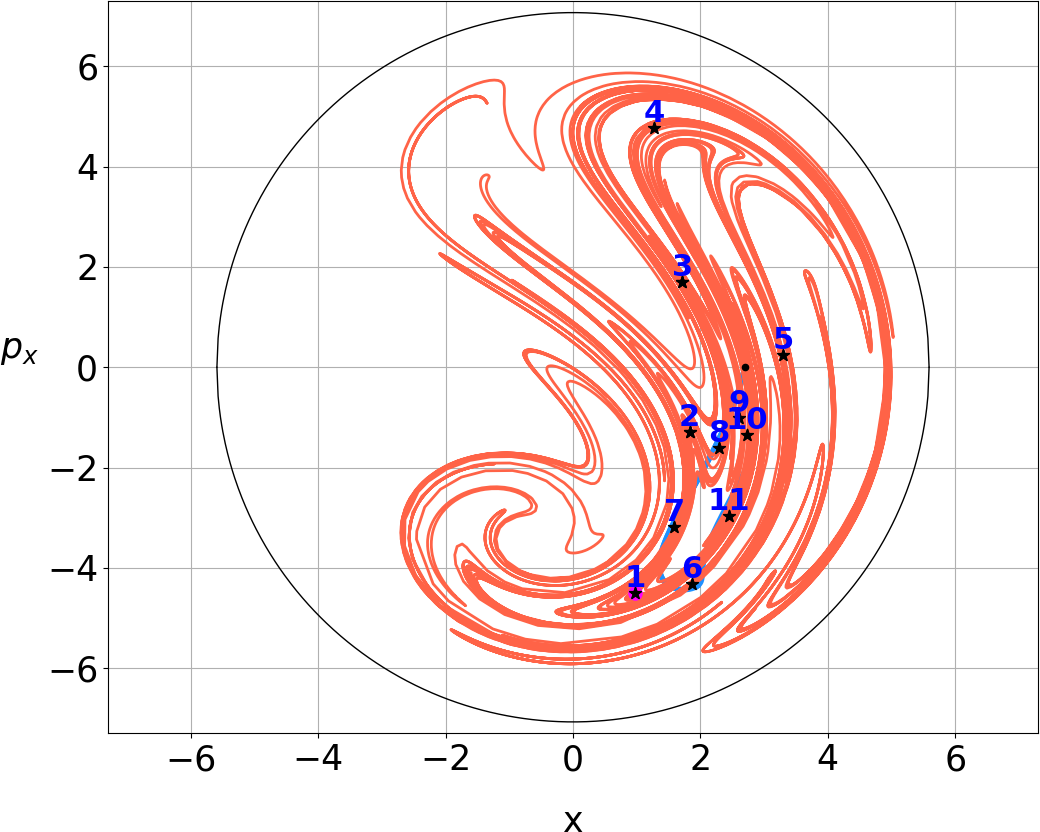}
\caption{The unstable and stable asymptotic curves up to $N=24$ in the case $h=25$ along with the points of the period-11 unstable periodic orbit. }\label{per11}
\end{figure}

 The first outer  lobe that enters $U_1'$ is $U_{11}'$, which has a part inside the inner lobe $U_6$ that enters $U_1'$. {In the common part of $U_{11}'$ and $U_1'$ we find a period-11 periodic orbit (Fig.~\ref{per11}), providing an explicit example of how the homoclinic lobe hierarchy generates high-period structures in the chaotic region.}
 
More generally for every value of the energy $h$ the index $i$ of the first outer lobe $U_i'$ that enter into the large chaotic sea is called ``first-penetration index'' and it is a function of $h$. This function is given in Fig.~6 and approaches a hyperobola. We see that $i(h)$ is relatively small for large $h$ but it increases considerably as $h$ decreases and tends to infinity as $h$ approaches a critical value $i_{crit}$ for $h=h_{crit}=23.3$.

{In fact, for $h<h_{crit}$ the invariant curves around $O$ prevent the asymptotic curves of $O$ from reaching the outer chaotic zone produced by the unstable orbit $O_1O_2O_3$. For $h>h_{crit}$ the invariant curves no longer provide a complete barrier and the asymptotic curves from $O$ and $O_1O_2O_3$ intersect at heteroclinic points \cite{Contopoulos200210}. Thus, the destruction of the last invariant barrier produces topological communication, but the increasingly large value of $i(h)$ shows that efficient global transport is strongly delayed near the threshold.}


\begin{figure}[H]
\centering
\includegraphics[scale=0.35]{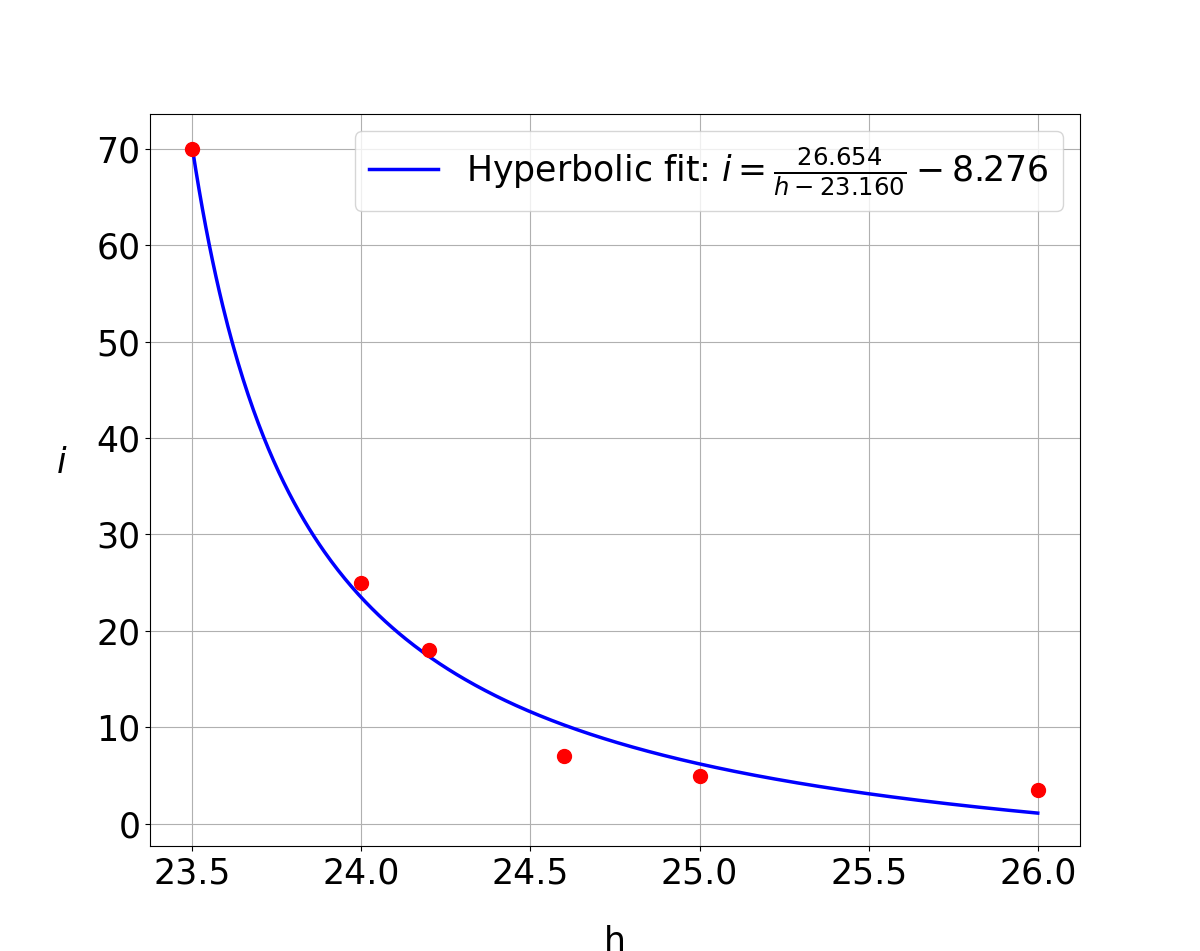}
\caption{The index $i$ of the first lobe $U'_i$ that  reaches the global chaotic sea, as a function of the energy $h$. The corresponding data (red dots) can be fitted by a hyperbola (blue curve).}\label{numlobes}
\end{figure}

\subsection{The case of the triple unstable periodic orbit $O_1O_2O_3$}

{If the ratio $\omega_1/\omega_2$  of the unperturbed frequencies is an irrational number, then, in the  Poincaré section $y=0$ with $p_y>0$, the unperturbed system has only the central periodic orbit $O$. In the perturbed system, however, there are several sets of stable and unstable periodic orbits. The asymptotic curves from some unstable periodic orbits intersect, producing chaos. This happens when the invariant curves  separating two different periodic orbits are destroyed and become cantori. And as the energy $h$ increases, the asymptotic curves of more distant periodic orbits intersect and the chaotic region increases considerably.}

{On the other hand, if $\omega_1/\omega_2=m/n$ is a rational number,  then the unperturbed motion is periodic. In the $y=0,p_y>0$ Poincar\'e section the resonance gives rise to periodic orbits of period $n$, since after one return to the section the phase of the first oscillator advances by $2\pi m/n$. Under a nonzero perturbation, the resonant periodic-orbit families are split into isolated stable and unstable periodic orbits. }

{All the results presented in this work correspond to the rational resonant case $\omega_1/\omega_2=4/3$. Thus, the relevant non-central resonant periodic orbits in the Poincaré section have period $3$. In particular the system contains a central  simple stable periodic orbit $O$ and a triple periodic orbit $(O_1O_2O_3)$, which is unstable and lies between the three surrounding stable period-$3$ islands.
The asymptotic curves of the triple unstable periodic orbit $O_1O_2O_3$ consist of two stable and two unstable curves from each of the points $O_1$, $O_2$, and $O_3$. Thus, the total number of asymptotic curves is $12$.}

\begin{figure}[H]
\centering
\includegraphics[scale=0.09]{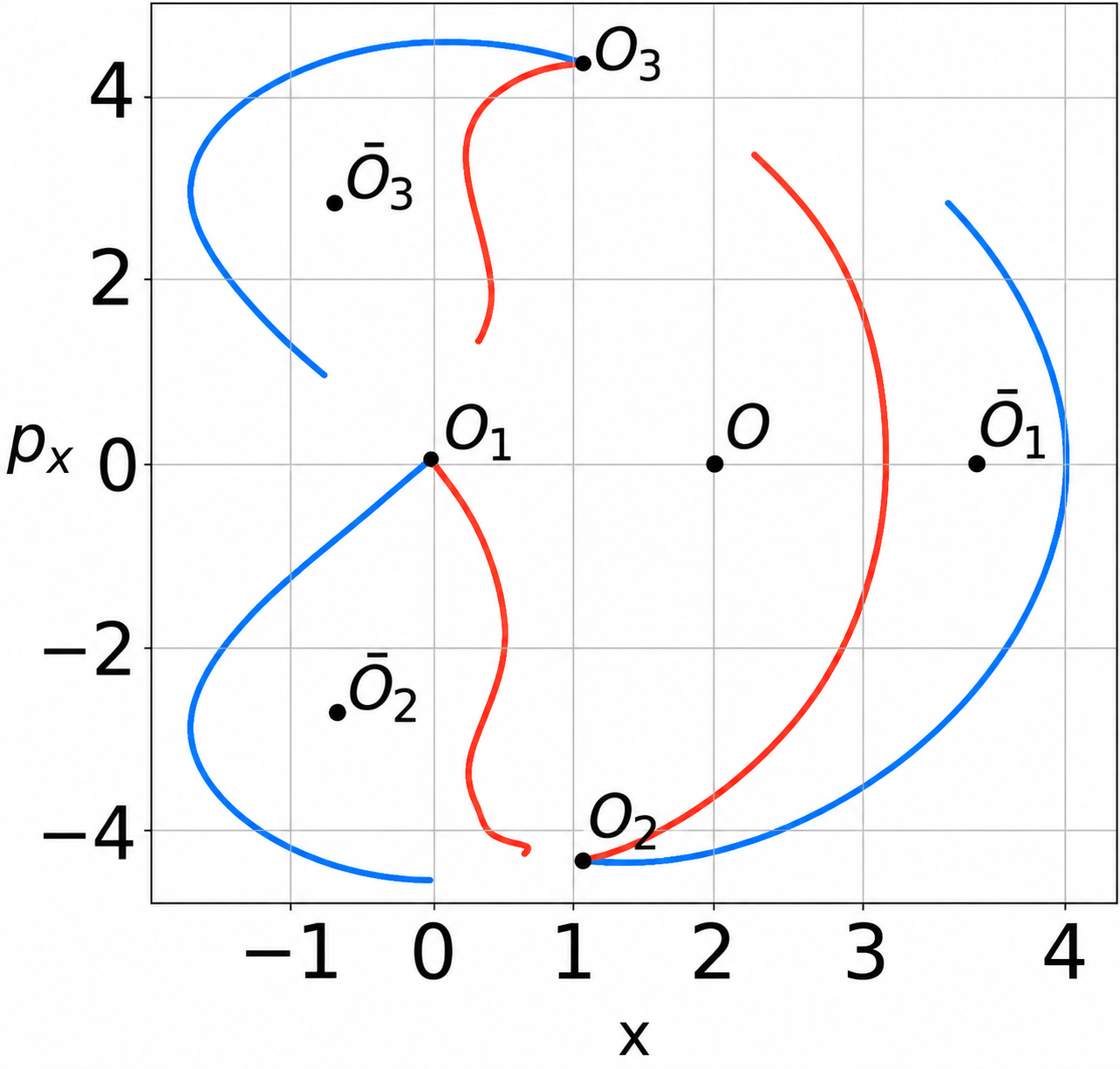}[a]\\
\includegraphics[scale=0.22]{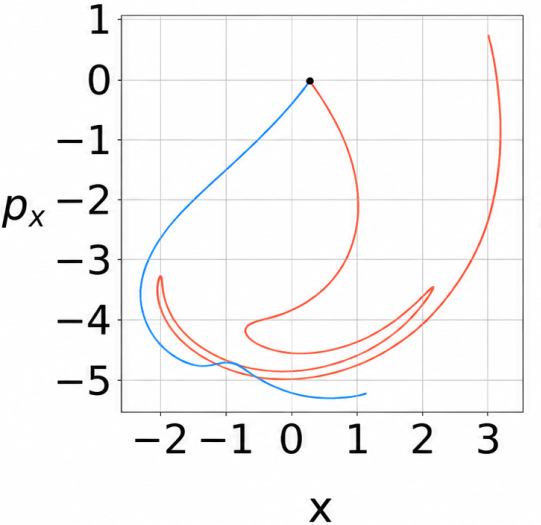}[b]
\includegraphics[scale=0.22]{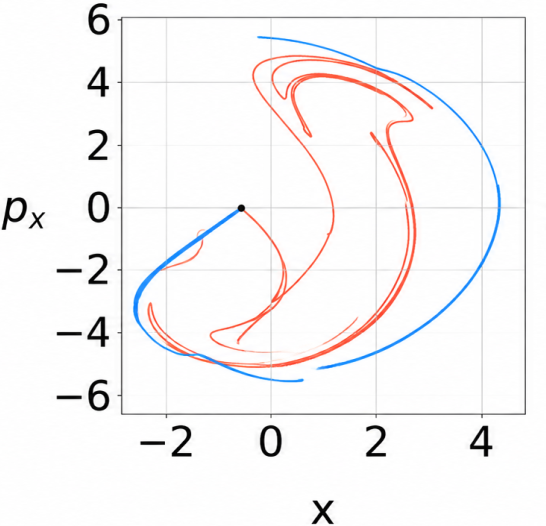}[c]
\includegraphics[scale=0.22]{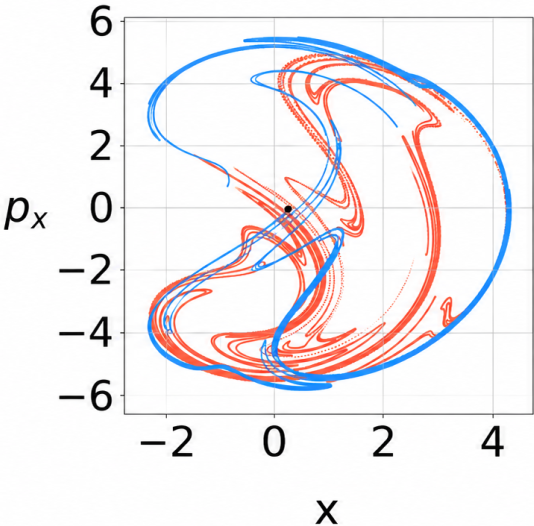}[d]
\caption{{a) Two asymptotic curves from  every point of the unstable periodic orbit $O_1O_2O_3$ for $h=19$ and $N=30$ iterations (stable blue and unstable red). The oscillations in this case are very weak. The asymptotic curves from  the point $O_1$ for $h=23.2$ and b) $N=18$ iterations, c) $N=21$ iterations and d)  $N=25$ iterations.}}\label{23comma2}
\end{figure}

{Figure~\ref{23comma2} illustrates the development of the asymptotic curves
of the triple unstable periodic orbit $O_1O_2O_3$. For $h=19$ and
$N=30$ (Fig.~\ref{23comma2}a) only small oscillations of the stable asymptotic 
curves are visible. At $h=23.2$, the oscillations become much more
pronounced as the number of iterations increases. For $N=18$
(Fig.~\ref{23comma2}b), large oscillations of the stable curve and the
beginning of oscillations of the unstable curve appear near $O_2$.
For $N=21$ (Fig.~\ref{23comma2}c), the oscillations become more numerous
and form loops around $O_1$, $O_2$, and $O_3$. Finally, for $N=25$
(Fig.~\ref{23comma2}d), the manifold structure becomes still more
extended, with the curves originating from $O_1$ approaching the other
two points of the unstable periodic orbit. Nevertheless, the asymptotic
curves cannot approach the central orbit $O$, which remains surrounded
by invariant curves. In this case, the eigenvalues along the stable and
unstable directions are $\lambda_1=18.913$ and
$\lambda_2=1/\lambda_1=0.053$.}


\bigskip

\section{{Asymptotic Curves Beyond the Escape Energy}}

{A qualitatively different dynamical regime appears when part of the invariant-manifold network opens to infinity. We therefore examine a generalized H\'{e}non-Heiles Hamiltonian for which two of the four asymptotic branches of the main unstable periodic orbit extend to infinity. The important point is not only that escape becomes possible, but that the opening of two branches to infinity changes the topology of the remaining non-escaping manifold network: the branches that intersect no longer form the same enclosing lobe structure found in the bounded case.}

The problem considered above refers to perturbations of the form $xy^2$. For different perturbations we find, in general, similar phenomena.

In particular consider the generalized H\'{e}non-Heiles Hamiltonian \eqref{Ham} with $\varepsilon=1$
\begin{equation}\label{ham2}
H = \frac{1}{2}(p_x^2 + p_y^2 + \omega_1^2 x^2 + \omega_2^2 y^2)
+ xy^2 + \alpha x^3  = h,
\end{equation}
which is close to an integrable case.

The system \eqref{ham2} is integrable in three cases:
 (1) for $\omega_1 = \omega_2$, $\alpha = -\frac{1}{3}$,
 (2)  for arbitrary $\omega_1,\omega_2$ and $\alpha = 2$, and (3) for  $\omega_1 = 2\omega_2$ and $\alpha = \frac{1}{6}$ \cite{bountis1982integrable}.

We consider now the particular case when $\varepsilon=1$, $\omega_1 = \omega_2 = 1$ and $\alpha$ close to $\alpha=2$ for various values of the energy $h$. Then  from Eq.~\eqref{ham2} 
%
%
we derive  the zero velocity curve:
\begin{equation}
y^2 = \frac{2h - x^2 - 2\alpha x^3}{1 + 2x}.
\end{equation}
This tends to $y^2 \to \infty$ for $x \to -\tfrac{1}{2}$ and the numerator  $F=2h - x^2 - 2\alpha x^3$ is positive. The point $(x=-\frac{1}{2},p_x=0)$ is well to the left of the periodic orbit  for $\alpha$ close to $\alpha=2$. E.g. for $\alpha=1.5$  it is $(x=-0.264, p_x=0)$.

The numerator has either one or three roots and becomes minimum $F_{\min}$ if $dF/dx=0$, i.e. when
\begin{equation}
x_c = -\frac{1}{3\alpha}.
\end{equation}
The minimum value is at $p_x=0$  and consequently the escape energy reads:
\begin{equation}
h_{esc} = \frac{1}{2}\left(x^2 + 2\alpha x^3\right)
= \frac{1}{54\alpha^2}.
\end{equation}
For $\alpha=2$ we find $h_{esc}=0.0046$ and for $\alpha=1.5$ $h_{esc}=0.0082$.

For $h<h_{esc}$ the  orbits are inside a closed limiting curve given by Eq.~\eqref{lc}.



\begin{figure}[H]
\centering
\includegraphics[scale=0.166]{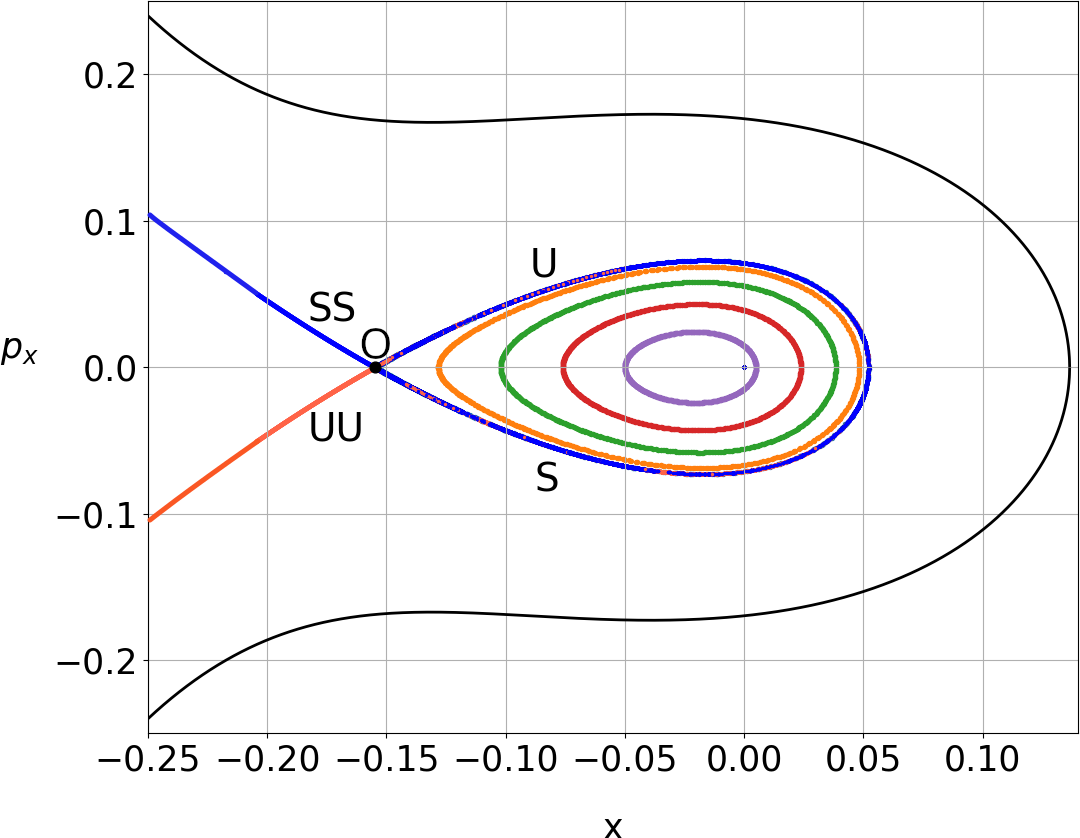}[a]
\includegraphics[scale=0.23]{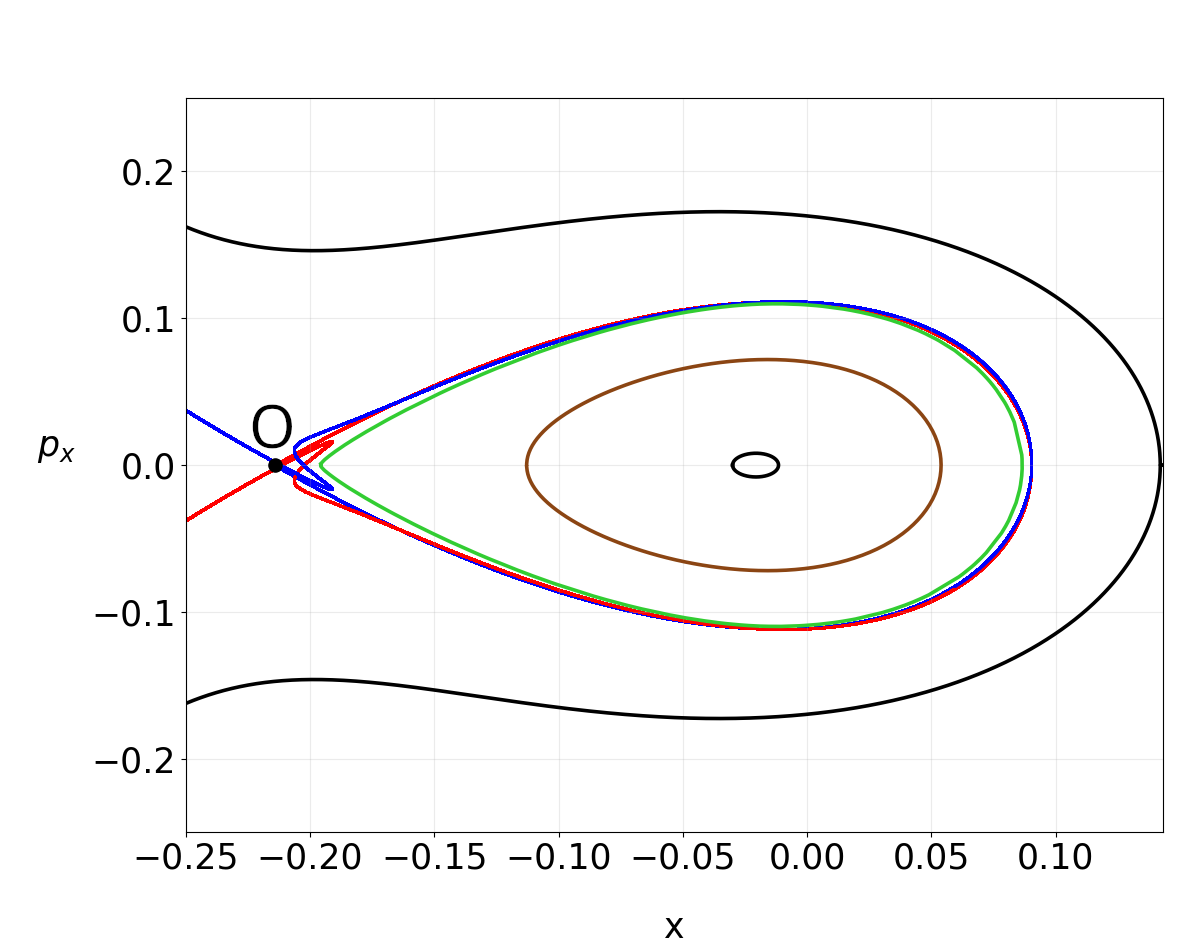}[b]
\caption{The asymptotic curves of the unstable periodic orbit $O$ and the surrounding limiting curves (black). a) In the case $\alpha=2$ the unstable and stable asymptotic curves on the right coincide. b) In the case $\alpha=1.5$ these curves intersect at homoclinic points.} \label{hen}
\end{figure}

{For $h>h_{\mathrm{esc}}$, an unstable periodic orbit exists on the surface of section at $p_x=0$. This orbit is characterised by a very strong instability, with largest eigenvalue $\lambda_1>2000$. Remarkably, this strong instability is also present in the integrable case $\alpha=2$ (Fig.~\ref{hen}a), where the invariant curves nevertheless remain well defined. The same holds also for $\alpha=1.5$ (Fig.~\ref{hen}b): the invariant curves also remain well defined, although the strong instability makes the numerical computation of the asymptotic curves to the right of $O$ particularly difficult (see the Appendix).}


From Eq.~6 we find on the surface of section $y=0$ for $\varepsilon=1$
\begin{equation}
p_y^2=2h-x^2-2\alpha x^3-p_x^2\geq 0.
\end{equation}
The boundary  $p_y^2=0$ (or $p_x^2=F=2h-x^2-2\alpha x^3$) is shown 
in Figs.~\ref{hen}a,b.  Outside this curve we have $p_y^2<0$ and thus  there are no real   orbits.

The case $\alpha=2$ has been studied in \cite{tzemos2025orbits} with $h=1$ for various values of $\varepsilon$.

In Fig.~\ref{hen}a we give the invariant curves on the surface of section $y=0$ for constant $\varepsilon=1$ and various values of $h$. The most important characteristic of this case  is that if $h$ is beyond the critical value $h=h_{esc}$,  there is an unstable periodic orbit on the surface of section $y=0$ (for $p_x=0$) that has two asymptotic curves on the right that join into one curve and two asymptotic curves on the left extending to infinity. This case has no chaos, but the orbits starting outside the closed asymptotic curve and on the left of O escape to infinity.

Then we consider cases with $\alpha$ close to 2, which have chaos but with a topology different from that studied in the  previous cases.

In these cases there is again an  unstable periodic orbit with two asymptotic curves going to infinity (Fig.~\ref{hen}b) but the other two asymptotic curves (stable and unstable) intersect each other  and do not join as in the integrable case.

The fact that two asymptotic curves on the left  extend to infinity does not  allow the other two  asymptotic curves to surround the unstable periodic orbit, as in the cases considered in section \ref{31}. This is why their topology is different.

\begin{figure}[H]
\centering
\includegraphics[scale=0.175]{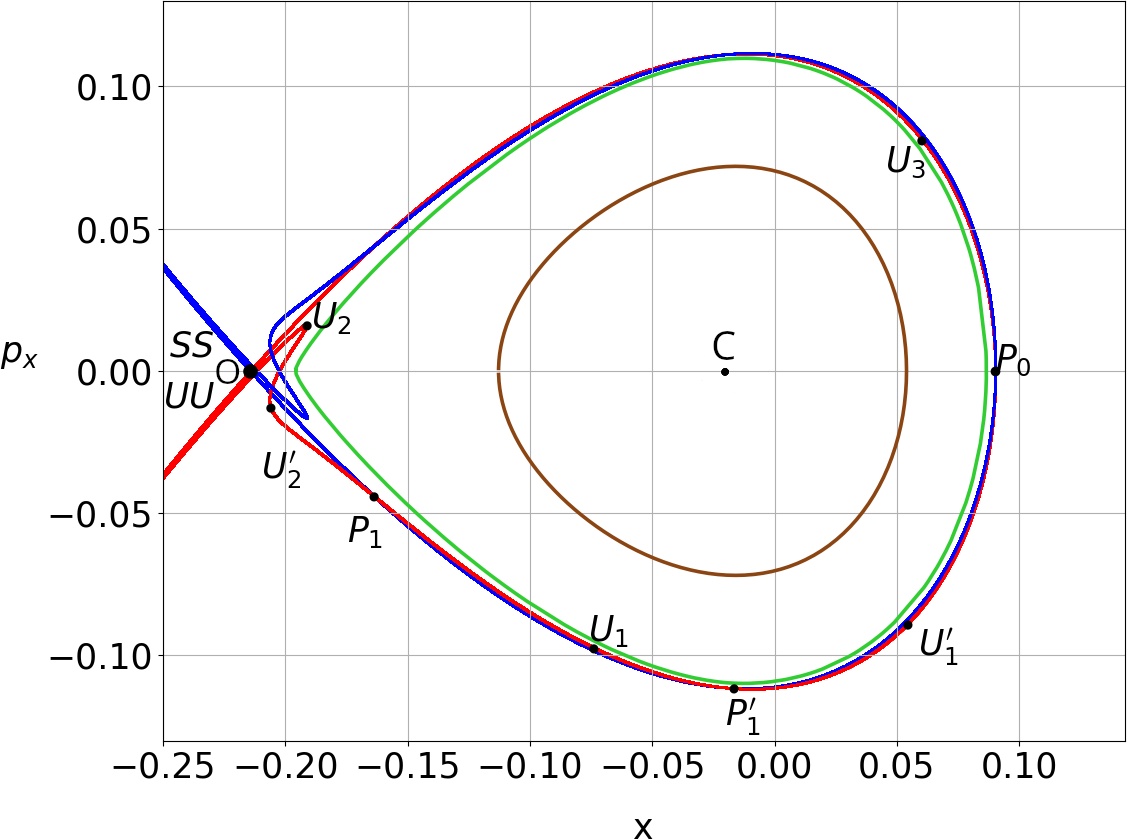}[a]
\includegraphics[scale=0.24]{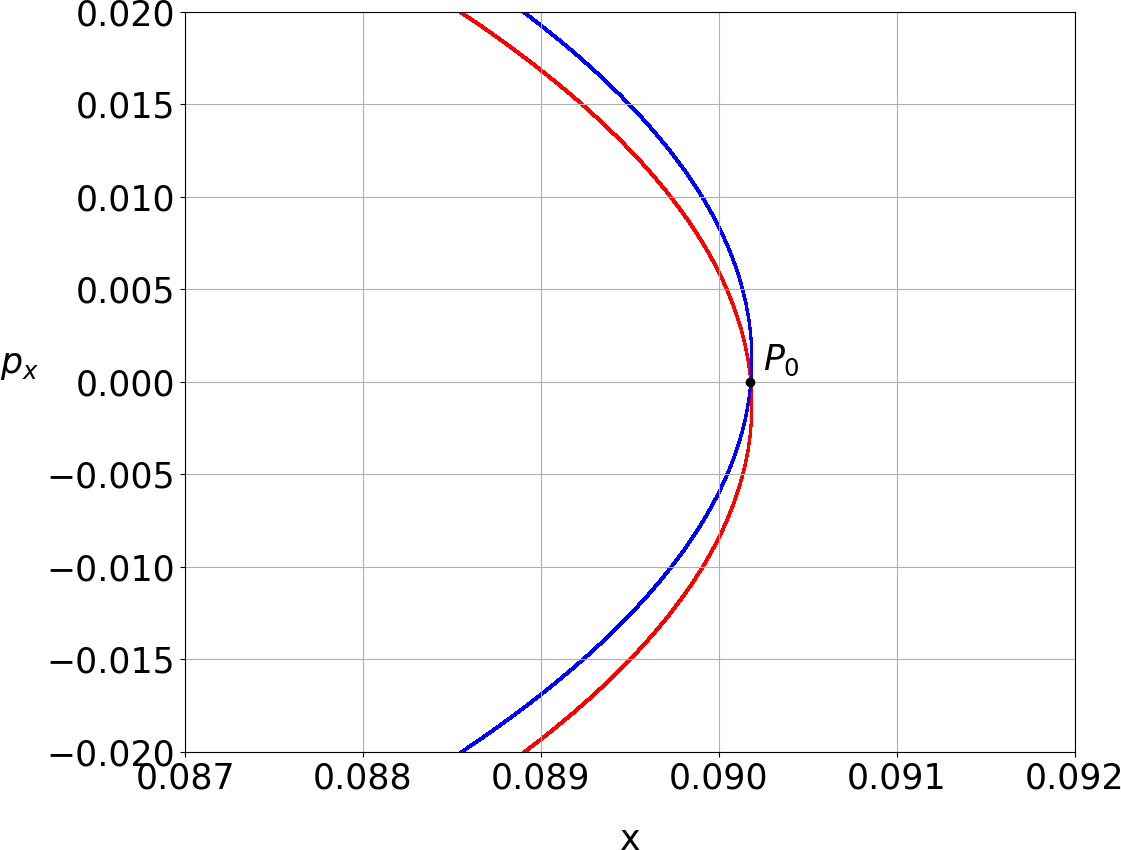}[b]
\includegraphics[scale=0.18]{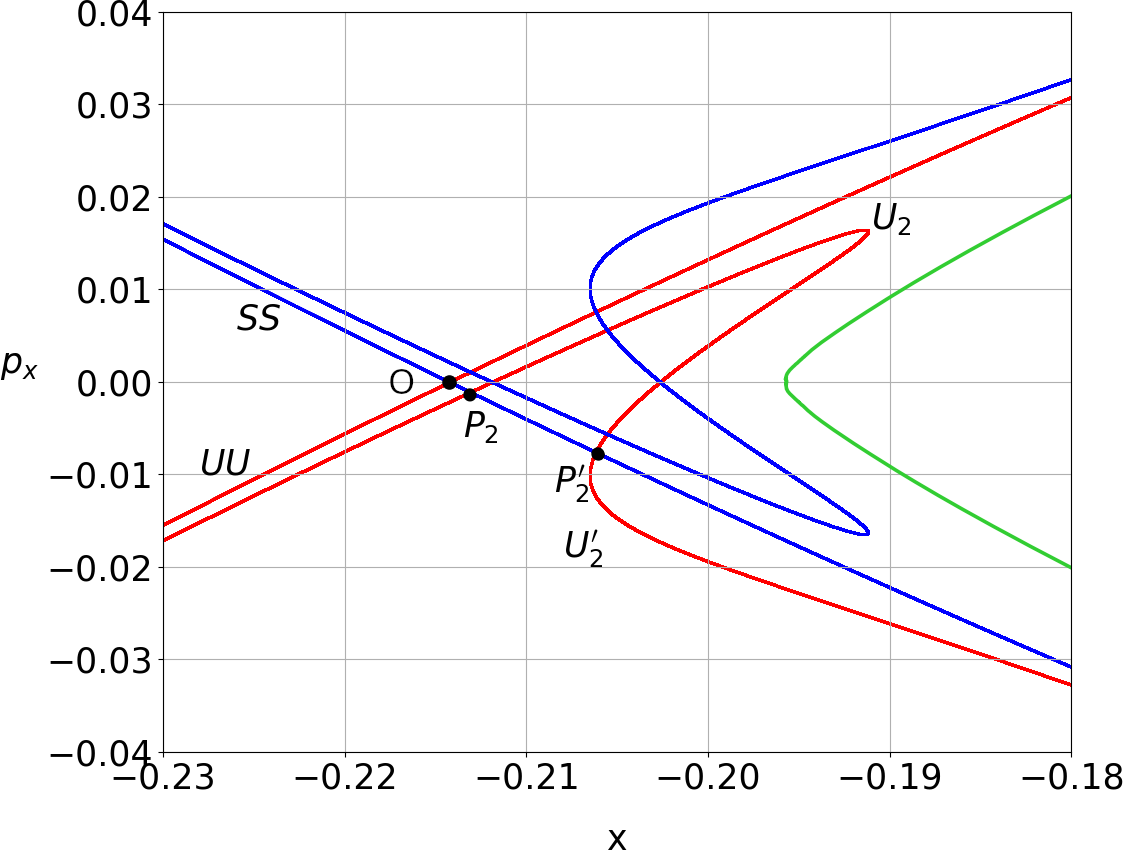}[c]
\caption{ a) The invariant curves of the unstable periodic orbit ({$\alpha=1.5, h=0.0144$})  in more detail  along with three typical KAM curves of the central stability island. b) zoom near the intersection point $P_0$ and c) zoom into the region near to $O$.} \label{hen2}
\end{figure}

In Fig.~\ref{hen2}a  we see on the right of the periodic orbit $O$ a set of closed invariant curves around the central periodic orbit $C$ ($x=-0.02065, p_x=0$) that are very similar to the closed invariant curves of Fig.~\ref{hen}a. The outermost closed  invariant  curve (green) starts at about $x=-0.196$ ($p_x=0$). But between the orbit $O$ at $(x=0.214, p_x=0$) and the green curve there is a chaotic region.

%

The orbits with initial conditions on the left of the point $O$, as well as the orbits close to the limiting curve in Fig.~\ref{hen}b, escape directly without further intersections with the surface of section. However, the periodic orbit $O$ generates asymptotic curves $U$ and $S$ that do not escape, but instead give rise to chaotic motion.

The asymptotic curves start along the four eigen-directions of the orbit $O$, with eigenvalues $\lambda_1=2.6081\times10^3$ and $\lambda_2=1/\lambda_1=3.8342\times10^{-4}$. The two branches on the left ($UU$ unstable, red, and $SS$ stable, blue) escape directly to infinity, whereas the other two branches  intersect repeatedly at homoclinic points.

The asymptotic curves $U$ and $S$ intersect for the first time at the point $P_0$ ($x=0.090176, p_x=0$) (we follow here the names of the  homoclinic points as in the case of the section \ref{31}). Beyond that point the outgoing invariant curve $U$ (red) goes downwards and to the left, just outside the curve $S$ (blue) (Figs.~\ref{hen2}a,b)  and  it crosses again the $S$ curve inwards at $P_1'$. Thus it forms the outer lobe $U_1'$ between $P_0$ and $P_1'$. Beyond $P_1'$ the $U$ curve proceeds to the left and crosses again the curve $S$ outwards at the point $P_1$ forming the inner lobe $U_1$ between $P_1'$ and $P_1$. Then it forms the outer lobe $U_2'$ up to the point $P_2'$ and the inner lobe $U_2$ up to the point $P_2$  (Fig.~\ref{hen2}c). It continues as a long outer lobe $U_3'$ very close to the asymptotic curve $UU$ on the left and,  next, a long inner lobe $U_3$ very close to the initial curve $U$ up to its tip $U_3$ (Fig.~\ref{hen2}a) before returning very close to $O$.

The next inner lobes are much longer and turn clockwise around the central periodic orbit $C$. The higher order lobes approach an ``asymptotic limiting curve'' around the origin very close to the green invariant curve of Fig.~\ref{hen2}a.  In fact, the ``asymptotic limiting curve'' coincides with the outermost invariant curve around the origin.

In the region between the orbit $O$ and the limiting invariant curve the orbits are chaotic. Most of them escape after a long time but some orbits like those starting on points of the stable asymptotic curve $S$ never escape. 
 
 The outer lobes are also very long close to $UU$  but we cannot compute them with a high accuracy (see Appendix). In any case the topology of the  outer asymptotic curves here is very different from that of the general cases described in Section~\ref{31}  because  in the present case these lobes cannot go around the periodic orbit $O$.

Similar phenomena have been found for other perturbations that have  periodic orbits beyond an escape energy  with  2 asymptotic curves extending to infinity (see details in \cite{Contopoulos200210,contopoulos1990asymptotic}).

\section{Conclusions}
{
We have studied the transition from local to global chaos in 2D
autonomous generalized H\'{e}non-Heiles systems, focusing on the role
of invariant manifolds in the development of chaos. We find that the destruction
of the last invariant curve leads to the  penetration of chaotic orbits into the
global chaotic region. However, the time required for this penetration may be very large.  The first-penetration lobe index $i(h)$ gives a
measure of this delay. The main results are as follows:}

\begin{enumerate}

\item {The chaotic fraction $P_{\rm ch}$ increases with energy through a
sequence of abrupt changes associated with the destruction of invariant
curves around resonant island chains. The resulting cantori allow
previously separated chaotic regions to communicate. Beyond the escape
energy $h_{\rm esc}$, $P_{\rm ch}$ also includes escaping orbits.}

\item {Before the destruction of the last invariant curve around an
unstable periodic orbit, the chaotic region around it remains separated
from the global chaotic sea. Above $h_{\rm crit}$, stable and unstable
asymptotic curves from different unstable periodic orbits intersect
heteroclinically, so that the two regions become connected. However,
the transport into the global chaotic sea is still slow near
$h_{\rm crit}$. This is shown by the rapid increase of the
first-penetration lobe index $i(h)$.}

\item {For the simple unstable periodic orbit $O$, the stable and unstable
asymptotic curves form inner and outer lobes. Their intersections also
produce high-period periodic orbits. In particular, a period-11 orbit is
found in the overlap of two lobes. The first outer lobe that reaches the
global chaotic sea occurs at an index $i(h)$ that increases strongly as
$h$ approaches $h_{\rm crit}$.}

\item {For the triple unstable periodic orbit $O_1O_2O_3$, the asymptotic
curves from one point of the orbit develop loops and lobes near the other
two points. Similar structures are also found for  other rational and
irrational frequency ratios.}

\item {A different topology is found above the escape energy when two of
the asymptotic branches extend to infinity. These branches do not take
part in the bounded lobe structure around the unstable periodic orbit.
The remaining branches form higher-order lobes around the central
invariant region and approach a limiting invariant curve.}

\end{enumerate}

{Overall, the results show that the increase of chaotic area with energy
is accompanied by changes in the topology of the invariant structures.
The first-penetration lobe index gives a way to describe the delay between
the destruction of the last invariant barrier and the penetration of the
global chaotic sea. The lobe structure also helps explain the appearance
of high-period periodic orbits.
}

{The conclusions above refer to the Hamiltonian forms, parameter ranges,
and frequency ratios studied here. Further work is needed to determine
how these results change for other perturbations and bifurcation
scenarios.}

\section*{Appendix}

The numerical construction of the unstable manifold for the periodic orbit in Fig.~\ref{hen2} is constrained by the strong instability of the system, reflected in the large leading eigenvalue $\lambda_1 \simeq 2608$. Initial attempts using very small segments (e.g., $\Delta L \le 10^{-10}$) led to severe numerical degradation, since exponential stretching under the Poincaré map increases distances between nearby points by several orders of magnitude per iteration. This leads to  a sparse sampling that cannot sustain a continuous representation, producing fragmented segments that fail to capture the global geometry. Moreover, initialization near machine precision ($\sim 10^{-16}$) amplifies round-off errors, which are exponentially enhanced and contaminate the dynamics with numerical noise.

However, the manifold remains nearly linear close to the hyperbolic fixed point, allowing a larger initial segment without violating the local approximation. We therefore use a dense initialization with $N = 5 \times 10^6$ points over $\Delta L = 10^{-6}$, remaining within the linear regime while improving robustness using the  LSODE integration scheme \cite{hindmarsh1983odepack} (spacing $\delta s \approx 2 \times 10^{-13}$). Although the initial conditions are iterated several times under the Poincaré map, the system’s extreme sensitivity implies that even tiny numerical errors in small regions of phase space may eject orbits from locally admissible regions, yielding incorrect dynamics. Consequently, long-time integrations become gradually unreliable.

\bibliographystyle{unsrt}
\bibliography{bibliography}

\end{document}